\documentclass{ws-ijmpd}

\begin{document}

\markboth{M.\ Spaans}
{A Bit of Quantum in GR}

%
\catchline{}{}{}{}{}
%

\title{Gravity and Information: Putting a Bit of Quantum into GR}

\author{M.\ Spaans}

\address{Kapteyn Astronomical Institute, University of Groningen\\ P.O. Box 800, 9700 AV Groningen, The Netherlands\\ spaans@astro.rug.nl}

\maketitle

\begin{abstract}
It is shown that quantum aspects of the energy-momentum tensor reveal dark matter and dark energy behavior for general relativity. Special meaning is assigned to the operator $X^2d^2/dX^2 +2Xd/dX$, for any field $X$ that is part of the energy-momentum tensor, and to the distinction between information on gravitational and non-gravitational effects. Changes in such information cause the Sun-Earth distance to increase by $11$ cm/yr.
\end{abstract}

\section{Quantum Aspects of the Energy-Momentum Tensor}

Einstein's famous elevator/rocket thought experiment unambiguously connects energy-momentum to space-time curvature as the expression of gravity, and leads to the Einstein equation[1]. The a piori separation of space-time and energy-momentum was not to Einstein's liking, despite the profound success of general relativity, and some deeper connection between the two may exist[1,2,3,4,5,6,7,8]. In this light, and as a follow-up to [5], quantum aspects are considered.

The energy momentum tensor is an observable quantity. Quantum-mechanically, this implies that it may be impacted by interaction with an observer. Indeed, the place, $s_1\equiv X\cdot$, and momentum, $s_2\equiv d/dX$, operators for some field $X$ in the energy-momentum tensor allow one to assign field values and the changes therein, but $s_1$ and $s_2$ do not commute. In fact, an observer that measures a field in the energy-momentum tensor (through $s_1$) also changes it (so acts with $s_2$). Through the Einstein equation this change propagates into space-time, which causes a secondary change (through $s_2$ again). It is this back-reaction that the observer should actually measure (through $s_1$ again). This leads to a ``measurement'' operator $M_X=s_1s_2s_2s_1=X^2d^2/dX^2 + 2Xd/dX$ that acts on any field in the energy-momentum tensor. The influence of the observer may seem small, {\it but} in reality any system observes itself unremittingly.

If implemented, $M_X$ would be a significant modification of Einstein gravity. Of course, one can argue that the measurement argument has no merit because the observer and observed parts of the gravitational system are assumed to be distinct entities in the first place. This very distinction, as an observable fact, implies that the action of the operator $M_X$ must be the identity. This is quite true. However, the distinction between observer and observed parts of the gravitational system can be made using {\it only} non-gravitational information. That is, irrespective of gravitational forces and, say, through exchange of information carried by light, the entanglement of an observer and a system part can be affected[9]. Consequently, $M_X$ does act if there is a difference between the non-gravitational and gravitational information that is necessary to describe the energy-momentum tensor. Therefore, one finds an operator $[M_X]^p$ with $p=0$ when a field $X$ is relevant to both gravitational and non-gravitational processes ($[M_X]^p$ is just the identity then) and $p=1$ when this is not the case.

Einstein gravity, for Newton's constant $G_N$, is given by ${\bf G}_{\mu\nu}/8\pi G_N {\bf T}_{\mu\nu}=1$ in the usual notation. Typically, $T^{\mu\nu}=(\rho +P/c^2) U^\mu \times U^\nu -P g^{\mu\nu}$, with density $\rho$, pressure $P$ and four-velocity $U^\mu$. Semi-classically, one can just transform a classical field $X$ in $T^{\mu\nu}$ to $[M_X]^pX$. E.g., under $M_X$ one has eigenvalues 2 and 6 in $M_P P=2P$ and $M_U U^2=6U^2$, yielding $M_{P,U} T^{\alpha\beta}=(\rho +2P/c^2) 6U^\alpha \times U^\beta -2P g^{\alpha\beta}$. This specific $p=1$ case pertains to a collisionless fluid, since in the absence of collisions no non-gravitational information on pressure and velocity is needed, and only density is required for the non-gravitational aspects of the system to be tangible. For pure dust the pressure is zero and one finds a factor 6 increase in the strength of gravity. Interestingly, this is equivalent to the 5 times more dark matter than baryonic matter that is needed to describe the dynamics of (almost collisionless) stellar systems like galaxies[5]. For systems that are partially collisionless (e.g., due to stellar collisions or gas hydrodynamics), the discrete variable $p$ varies with position depending on the local gravitational and non-gravitational information needed. Interestingly, below a redshift of $1-2$ fewer galaxies merge and most stars are in place, and the universe as a whole is relatively quiet. So more recently than redshift $1-2$ processes that need mostly gravitational information occur less frequently, by a factor of more than a few looking at the mass accumulation history of galaxies, compared to those that require gravitational and non-gravitational information (like the evolution of stars and diffuse matter). This favors $p=0$ for the global matter distribution of the universe since a redshift of $1-2$, compared to $p=1$ for earlier times, and leads to an acceleration in cosmic expansion (dark energy behavior).

\section{From Fields to Particles: the Sun-Earth System}

The operator $[M_X]^p$, when applied to individual particle excitations of an underlying field like density, should act exactly the same and weigh deviations in the required gravitational versus non-gravitational information of contributing particles. The Sun is obviously a collisionally dominated system and $[M_X]^p=1$. However, through particle interactions, the fusion cycle that powers the Sun causes our star to loose mass and $G_Nm(t)=G_Nm_0-G_N\delta m(t)+(f-1)G_N\delta m(t)$ for a starting mass $m_0$ in the classical limit. The $f$ term represents any extra gravitational information that is needed and follows from $T^{\mu\nu}$. For negligible pressure, $f=12$, i.e., 2 from $\rho$ times 6 from $U^2$ because now the non-gravitational information on {\it changes} due to the fusion cycle resides with individual particles only, independent of their mass density and four-velocity. In fact, one expects $f=36$ since 3 particles are removed for every produced $^4$He in the dominant fusion cycle of the Sun. This extra factor three is a pure particle effect not captured by the field description of $[M_X]^p$. In all, this translates into a correction of $34G_N\delta m(t)$ to $G_Nm_0$, which can be viewed as an information dependent change in the gravitational constant[5]. Therefore, the astronomical unit AU increases by $\sim 3\times 10^{-12}/yr$ due to the astronomical definition $G_Nm_\odot \equiv k^2 AU^3$[10], with $k$ Gauss' constant. This yields $\sim 11$ cm/yr, close to what appears to be observed[10].

\end{document}